\documentstyle[preprint,prb,aps]{revtex}

\def\AmS{{\protect\the\textfont2
        A\kern-.1667em\lower.5ex\hbox{M}\kern-.125emS}}

\makeatletter
\def\thepage{1-\@arabic\c@page}
\def\@pnumwidth{2em}
\makeatother
\begin{document}
\draft

\title{Magnetization under High Pressure in MnSi}

\author{Christophe Thessieu$^{\dag}$ , Kenji Kamishima$^{\ddag}$,
Tsuneaki Goto $^{\ddag}$} 
\address{$^{\dag}$ Department of Material Physics, Faculty of Engineering Science,\\
Osaka University, Toyonaka, Osaka 560}
\address{$^{\ddag}$ Institute for Solid State Physics, University of Tokyo, Roppongi,
Tokyo 106}
\author{G\'erard Lapertot}
\address{Centre d'\'{E}tudes Nucl\'{e}aires de Grenoble, SPSMS, Grenoble, France}
\date{\today}
\maketitle

\begin{abstract}
The magnetization M(H) has been measured in the weakly helimagnetic
itinerant compound MnSi under high pressure up to 10.2~kbar and high
magnetic field up to 9~Tesla. We interpret the simultaneous decrease under
pressure of the saturated magnetization, $p_{\text {s}}$, and the Curie temperature, $%
T_{\text{c}}$ in the frame of the self-consistent renormalization theory (SCR) of
spin fluctuations. From the analysis of the so-called Arrot-plot
($H/p [ H,T ] $ versus $p^2[ H,T ] $) and the
respective volume dependence of
$p_{\text{s}}$ and $T_{\text {c}}$, we estimate the
evolution of the characteristic spin fluctuation temperatures,
$T_0$ and $T_{\text{A}}$ when the system approaches its critical pressure,
$P_{\text {c}}$=15~kbar, corresponding to the disappearance of the long
range magnetic order at $T=0$.
\end{abstract}


\section{Introduction}

MnSi is a well-known weakly helimagnetic system which has given a lot of experimental
supports to the Self Consistent Renormalized spin fluctuations theory
(referred henceforth as the SCR theory) developed by Moriya and co-workers
(for a general review see Ref.\ \onlinecite{MOR85}). At ambient pressure both
macroscopic (electrical resistivity, magnetic susceptibility,
magnetization)\cite{FAW70,KAD82} and microscopic properties
(nuclear magnetic relaxation and neutrons
scattering) \cite{YAS78,ISH85} are quantitatively well interpreted
in the frame of a SCR treatment of the magnetic spin fluctuations. Although
the SCR theory furnishs
a phenomenological treatment of the long wavelength-low energy
spin-fluctuations, it successfully explains the low value of the
Curie temperature $T_{\text{c}}$=29.4~K, the $T^2$ low temperature thermal variation
of the electrical resistivity $\Delta \rho =AT^2$, the high temperature
constant value exhibited by the spin-lattice relaxation rate $T_1$, and
above all the puzzling Curie-Weiss like behaviour of the paramagnetic
susceptibility, $\chi^{-1}(T)\propto (T-T_{\text{c}})$. Magnetization
measurements under pressure in this compound
have already been reported in a previous study up
to 5.2~kbar by Bloch et Al.\cite{BLO75}. However since that time several
studies have shown the crucial role played by the hydrostatic pressure on
the magnetic ground state of this compound \cite{PFL94,PFL95,THE95}.
In particular, the existence of a critical pressure, $P_{\text{c}}$=15~kbar,
where the system undergoes a magnetic to non magnetic transition
at $T=0$ (Quantum Phase Transition) has been pointed out\cite{PFL94,THE95}.
Recently, in the vicinity of this quantum critical
point, it has been interestingly sheded
light on the fact that the disappearance of the long range helimagnetic order was
accompanied by the occurrence of the so-called \textit{Itinerant Electron
Metamagnetic} transition under an external magnetic field\cite{THE96}. This
first order magnetic phase transition is a common feature shown by other 3d itinerant
magnetic systems, especially when the external temperature independent
parameter tuning the quantum magnetic instability
is the chemical substitution, $x$, (Y(Co$_{\text{1-x}}$Al$_{\text{x}})_2$
\cite{SAK90}, Co(Se$_{\text{1-x}}$S$_{\text{x}}$)$_2$
\cite{ADA79}) and more seldom the hydrostatic pressure, $P$, (CoS$_2$ \cite{END96}).
Although this magnetic phenomena has been largely enligthened by various
experimental works, its theoretical treatment is actually not unanimously
recognized with notably the ambiguity between the role played by the band
structure effect at the Fermi level\cite{YAM93} and the influence of an
external parameter on the spin fluctuations spectrum \cite{TAK95}. In view
of this uncleared situation, it appears interesting to extend the measurements
of the magnetization properties in this compound up to the highest pressure
actually available with our technique (11~kbar).

\section{Experimental Details}

For this experiment a single crystal of MnSi with parallelepiped shapes and
typical dimensions of 3$\times$0.3$\times$0.5~mm (total weight of 29.4~mg)
has been measured. The measurements were performed using an He$^4$ cryostat in a
temperature range from 4.2~K to 60~K. The external magnetic field was
supplied by a 9~Tesla superconducting magnet and the weak magnetic anisotropy
of MnSi has permitted to apply the external field in an undefined
crystallographic direction. The hydrostatic pressure was generated by a
non-magnetic clamp-type pressure cell of small dimensions with a maximal enable
pressure of almost 11~kbar. The original Ti-Cu alloyed composition of the
clamp core added to a set of ceramic-type pistons give a small diamagnetic
contribution to the background signal allowing absolute value measurements
under pressure even in a strong external magnetic field. The
principle of the measure is based on the oscillation of the whole pressure cell in a two
pick-up coils system. The signal is analyzed by the extraction method. This
experiment has been carried out at the Institute for Solid State Physics in
the University of Tokyo.

\section{Results}

The Fig.\ \ref{fig1} gives the evolution of the isothermal magnetization process at
$T=4.2$~K, for magnetic field up to 9~Tesla and under different applied
pressures. At ambient pressure the magnetic features are well understood as
a rotation of the helical structure along the field direction
($H_1\sim 1$~kOe) followed by the disappearance of the magnetic domains in an unique
conical phase (1~kOe$<H<$6.2~kOe) orientated in the field-direction.
Above $H_2=6.2$~kOe, where the abrupt kink observed in the curve $p[H]$
corresponds to the transition in an induced ferromagnetic state,
we observe an unsaturated ferromagnetic behavior. All
these features remain unchanged under pressure and it is noteworthy that
the value of the characteristic fields, $H_1$ and $H_2$, remain unchanged
under pressure. The insert of the Fig.\ \ref{fig1} shows the magnetization behavior at
$P=10.2$~kbar for the both way of sweeping the magnetic field. Below 1~kOe a
tiny hysteresis is observable and clearly in this region the magnetization
does not respond linearly to the applied field. However above 1~kOe this
hysteresis disappears and the magnetization corresponds to the conical phase
and the unsaturated behavior of the induced ferromagnetic state above $H_2$.
These facts ensured a tangible proof that the heli-magnetic structure of
MnSi persists even under high pressure (at least up to 10.2~kbar) and that
no new magnetic structure transition is induced by pressure. By intercepting
the high field regime of the magnetization at $H=0$, we obtain the value of
the saturated magnetization $p_{\text{s}}$, which at ambient pressure exhibits
a value of 0.39~$\mu _{\text{B}}$/Mn in accordance with previous results\cite{BLO75}.
From the Fig.\ \ref{fig1}, we see that the pressure effect is to lower the
value of the magnetization and in particular leads to a decrease of the
saturated magnetization value at
$T=4.2$~K, $p_{\text{s}}[H=0, 4.2~\text {K}]$. In
order to extract a precise quantitative variation of
$p_{\text{s}}[P]$ we have
presented on the Fig.\ \ref{fig2} the so-called Arrot-plot $H/p[H, T ]$ versus
$p^2[H, T]$ which gives a direct way to obtain this value.
The axis have been drawn in a reduced units system such
as the external magnetic field is expressed in energy units
$h=2\mu _{\text{B}}H$ and the magnetization per magnetic atom
($p=2\frac{M(T)}{N_0}$, $N_0$ number of magnetic sites),
is expressed in $2\mu _{\text{B}}$ units. To obtain the saturated
magnetization $p_{\text{s}}$ for each pressure we have estimated the
$x$-axis intercept of the high magnetic field linear extrapolation.
The evolution of $p_{\text{s}}$, reported in the insert of the Fig.\ \ref{fig2}, is
surprisingly weak with pressure and this contrasts with the strong decrease
of the corresponding Curie temperature $T_{\text{c}}$. It is noteworthy to
underline that even close to the critical boundary ($P_{\text{c}}=14.8$~kbar)
the saturated magnetization still exhibits a high value.
Quantitatively under an applied pressure of 10.2~kbar, the value of the
saturated magnetization is only modified from 0.39 to
0.34~$\mu _{\text{B}}$/Mn atom whereas $T_{\text{c}}$
drops from 29.4~K to 13.4~K (cf. table\ \ref{table1}
for the detailed values). The weak decrease of the magnetization has been
confirmed recently by zero-field NMR measurements under pressure\cite{THE97}
showing the weak decrease of the transferred hyperfine field of the Mn sites
with pressure. In a good agreement with previous authors \cite{BLO75} we
find that
$d\ln p_{\text{s}}/dP=-1.27 \: 10^{-2}\text{kbar}^{-1}$
($-1.15 \: 10^{-2}\text{kbar}^{-1}$
in the case the previous results of Bloch et Al.). The Fig.\ \ref{fig3} shows the
isothermal $p[H]$ curves around $T_{\text{c}}$ for a pressure of 10.2~kbar.
No metamagnetic transition either hysteretic process are observed around
$T_{\text{c}}$, in accordance with the fact that the transition is still of a
second order nature up to 12.5~kbar\cite{PFL95}.

\section{Discussion}

In the frame of the SCR theory, the magnetic spin fluctuations are treated in a self
consistent way and integrated in the magnetic equation of state. The nature
of the spin fluctuations is therefore characterized by a small set of
parameters expressed as $p_{\text{s}}$, $\bar{F_1}$, $T_0$ and $T_{\text{A}}$.
The two first parameters are
determinated by the knowledge of the magnetic equation of state given by the
Landau development of the magnetic free energy in the presence of an
external magnetic field
\begin{equation}
-\frac{2(\alpha-1)}{\rho}+\bar{F_1}p^2=\frac{2\mu_{\text{B}}H}{p}
\label{equ1}
\end{equation}
where $\alpha=I\rho$, $I$ is the intra-site coulombian interaction, $\rho$ the
density of states at the Fermi level, $p$ the magnetization expressed in
$2\mu_ {\text{B}}$ units (the saturated magnetization at $T=0$, $p_{\text{s}}$ is with this
convention such as $p_{\text{s}}=2M(0)/N_0$) and the parameter $\bar{F_1}$ is the
coefficient of mode-mode coupling between magnetic spin fluctuations.
Consequently, it is possible upon magnetization measurements and the
analysis of the Arrot-plot defined with our units systems such as
$2\mu _{\text{B}}H/p[H,T]$
versus $p^2[H,T]$, to obtain
quantitative values for these two parameters. The two remaining parameters
$T_0$ and $T_{\text{A}}$ are directly related to the imaginary part of the generalized
dynamic susceptibility Im$\chi(q, \omega)$. Neutrons measurements on MnSi
have clearly shown that in the weakly ferromagnetic limit, Im$\chi(q, \omega)$
has the lorentzian form in a small region of $q$ and $\omega$ close to the
ferromagnetic instability vector $q=0$ and is expressed as: 
\begin{equation}
Im\chi(q, \omega)=\frac{\chi (0,0)}{1+\frac{q^2}{\kappa^2}}\times
\frac{\omega\Gamma_q}{\omega^2+\Gamma_q^2}
\label{equ2}
\end{equation}
with $\chi(0,0)$ the static uniform susceptibility, $\Gamma_q$ the spectral
width of the magnetic SF excitations given by: 
\begin{equation}
\Gamma_q=\Gamma_0q(\kappa^2+q^2)
\label{equ3}
\end{equation}
$\kappa$ is the inverse of the magnetic correlation length. This power
spectrum of the magnetic spin fluctuations, Im$\chi(q, \omega)/\omega$ is
in fact characterized by two temperatures scales, $T_0$ and $T_{\text{A}}$,
representing respectively the dispersion along the $\omega$-axis and the
$q$-axis
\begin{eqnarray}
T_0&=&\Gamma _0 q_{\text{B}}^3/2\pi \nonumber \\
T_{\text{A}}&=&\frac{q_{\text{B}}^2}{2\kappa^2} \frac{N_0}{\chi(0)} 
\label{equ4}
\end{eqnarray}
The effective boundary vector $q_{\text{B}}$ is given by
$q_{\text{B}}=(6\pi^2/v_0)^{1/3}$
($v_0$ the volume per magnetic atoms). Usually these two temperatures scales are
obtained upon microscopic measurements such as nuclear magnetic resonance
(spin-lattice relaxation rate and Knight shift) for $T_0$ and neutrons
diffusion measurements for $T_{\text{A}}$. The SCR theory \cite{TAK86} predicts a
number of interesting crossed-relations between these differents parameters. For
instance, the mode-mode coupling term can be remarquably expressed as
function of the two temperatures scales, $T_0$ and $T_{\text{A}}$
\begin{equation}
\frac{\bar{F_1}}{k_{\text{B}}}=\frac{4}{15}\times \frac{T_{\text{A}}^2}{T_0}
\label{equ5}
\end{equation}
$T_0$ and $T_{\text{A}}$ can also be expressed in
the following way as function of
$T_{\text{c}}$, $p_{\text{s}}$ and the ratio $\bar{F_1}/k_{\text{B}}$ 
\begin{eqnarray}
T_0^{5/6}&=&\frac{10.334}{p_{\text{s}}^2}
\left( \frac{k_{\text{B}}}{\bar{F_1}} \right) ^{1/2}T_{\text{c}}^{4/3} \nonumber \\
T_{\text{A}}^2&=&\frac{38.737}{p_{\text{s}}^2}
\left( \frac{\bar{F_1}}{k_{\text{B}}}\right)^{1/2}T_{\text{c}}^{4/3}
\label{equ6}
\end{eqnarray}
From the following expressions, we readily see that upon macroscopic
measurements such as magnetization and the Arrot Plot analysis, giving
access to the ratio $\bar{F_1}/k_{\text{B}}$, $p_{\text{s}}$ and the knowledge of the
ordering temperature $T_{\text{c}}$ \cite{PFL94,THE95},
we can estimate quantitatively the values of the
energy scale of the spin fluctuations spectrum. The different values obtained
upon the Arrot-plot analysis and the Eqs.\ (\ref{equ4}) ($\Gamma_0$) and (\ref{equ6}) ($T_0$
and $T_{\text{A}}$) are listed
in the table\ \ref{table1}.
In the Fig.\ \ref{fig4}, we have compared the pressure effect on the characteristic temperatures
in MnSi, $T_{\text{c}}$, $T_0$ and $T_{\text{A}}$. As the pressure increases and the system is driven
towards its quantum critical point, the energies $T_0$ and $T_{\text{A}}$ change
in an appreciable way, indicating that
the spectra of the magnetic excitations is largely modified
by pressure due to a strong volume dependence of Im$\chi (q, \omega)$ in
this compound. Consequently,
we can suppose that from both sides of the critical pressure, the spatial
distribution of Im$\chi(q, \omega)$ is drastically modified and that a
discontinous variation of $T_0$ and $T_{\text{A}}$ is expected, corresponding to a
modification of the magnetic spin fluctuations regime when crossing
the critical boundary from the electronic spin-polarized state
to the paramagnetic Fermi liquid.
In particular the sizeable decrease of the energy dispersion
of the spin fluctuations modes, $\Gamma _0$, presented in the Fig.\ \ref{fig5} indicates that
these latter are more drastically over-damped at the approach of $P_{\text{c}}$
where the long range magnetic order has been shown to disappear and the
system remains paramagnetic at all temperature. Upon our
experimental results we can estimate that the pressure effect on the damping
rate of the magnetic spin fluctuations should be in the order of 
\begin{equation}
\mathrm{\frac{\partial \ln \Gamma _0}{\partial P}=-8.6 \: kbar^{-1}}
\label{equ7}
\end{equation}
On the other hand, the slope of the Arrot-plot is almost not modified by
pressure (Fig.\ \ref{fig2}) indicating that $\bar{F}_1$, the mode-mode
coupling term, has only a weak pressure dependence. We deduce therefore
that the
pressure modified the spin-fluctuations spectra distribution but not the
coupling strength between these latter. This scenario is
based on thermodynamic measurements and must be completed by
microscopic measurements under pressure. However, in view of these experimental
results, it appears important to verify the accuracy of this analysis by
additional measurements like neutrons inelastic diffusion experiment to
study the evolution of $T_0$ or $T_{\text{A}}$ under pressure.

\section{Conclusion}

The analysis presented in this paper tends to show that the evolution of the
spin fluctuations spectra with pressure plays a key role in the occurrence
of the Itinerant Electron Metamagnetic phenomenon as advanced recently by
Takahashi and Sakai. However the band structure effects induced by pressure
can not be readily analyzed from this measurement. Thus we do not exclude
that such effects at the Fermi level, as postulated by Yamada, are not engaged
in the Itinerant Electron Metamagnetism process. Our work mainly points out
that probably a more refined theoretical treatment is needed in order to
take into account the effects of both thermal and zero-point spin
fluctuations modes on the magnetic ground state of the system. It seems to
us that in MnSi the combined facts that there exists a well-pronounced
difference in the volume dependence
of the ordering temperature, $T_{\text{c}}$, and the saturated magnetization, $p_{\text{s}}$,
and the fact that in this itinerant electron system the phenomenon of metamagnetism has been observed
are strongly related.
We believed that the physical origin is due to a strong volume dependence of the
magnetic spin fluctuations spectrum in particular close to $P_{\text{c}}$.

One of us (C.T) is particularly indebted to both Prof.~Y.~Takahashi and
Prof.~H.~Yamada for fruitful and enlightening discussions about their
different standpoints. This work has been financially supported by a grant
of the European Union and the Japanese Society for the Promotion of Science.

\begin{figure}
\caption{Isothermal magnetization process at $T=4.2$~K under
different applied pressures. For each experiments, the pressure cell and the
sample have been cooled down in zero-field. The insert shows the
magnetization process at the maximal pressure obtained in this experiment
$P=10.2$~kbar close to $T_{\text{c}}$. The non-linearity of the
magnetization at low field ($H\simeq1$~kOe) is attributed to the persistence
of the helical magnetic structure at $P=10.2$~kbar.}
\label{fig1}
\end{figure}

\begin{figure}
\caption{Typical Arrot plot $h/p-p^2$ at $T=4.2$~K. The
intercept of the high field linear-fit with the x axis gives a direct value
of the saturated magnetization $p_{\text{s}}$. The evolution of
$p_{\text{s}}$ as function of the external pressure is reported in the insert and
can be compared with the pressure dependence of the ordering temperature $T_{\text{c}}$}
\label{fig2}
\end{figure}

\begin{figure}
\caption{Magnetization process for the maximal pressure
obtained in this study ($10.2$~kbar). For temperatures around
$T_{\text{c}}=13.4$~K, no metamagnetic either
hysteresis phenomena have been observed. All
the curves are presented for both way of sweeping the magnetic field.}
\label{fig3}
\end{figure}

\begin{figure}
\caption{Pressure dependence of the characteristic
temperatures $T_0$ and $T_{\text{A}}$ related to the imaginary part of the
generalized susceptibility Im$\chi(q, \omega)$. The error bars are mainly
due to the fitting error in the high field regime of the Arrot-plot.}
\label{fig4}
\end{figure}

\begin{figure}
\caption{The damping rate of the magnetic spin fluctuations $%
\Gamma_0$ shows a strong decrease in the vicinity of the critical boundary 
($P_{\text{c}}=14.8$~kbar). The error bars are calculated from the error on
$T_0$.}
\label{fig5}
\end{figure}

\begin{table}
\caption{Characteristic temperatures $T_0$ and $T_{\text{A}}$ of the
generalized susceptibility Im$\chi(q, \omega)$ calculated upon the
macroscopic measurements $p[H,P]$}
\label{table1}
\begin{tabular}{dcccccc}
Pressure[kbar]&$T_{\text{c}}$~[K]&
$p_{\text{s}} [ \mu _{\text{B}}\text{/Mn}]$&
$\bar{F}_1/k_{\text{B}}$~[K]&$T_0$[K]&
$T_{\text{A}}$[K]&$\Gamma_0$[$k_{\text{B}}$\AA$^3$] \\ \tableline
Ambient & 29.07 & 0.389 & 9055 & 146 & 2230 & 292 \\ 
3.6 & 24.01 & 0.374 & 9314 & 117 & 2025 & 234 \\ 
6.6 & 19.60 & 0.359 & 8899 & 97 & 1804 & 195 \\
8.7 & 16.27 & 0.348 & 8507 & 78 & 1576 & 155 \\
10.2 & 13.50 & 0.342 & 8401 & 60 & 1384 & 121\\ 
\end{tabular}
\end{table}

\end{document}